\documentclass[a4paper]{article}

\usepackage[inner=1in,outer=1in,top=1in,bottom=1in]{geometry}
\usepackage{microtype}                 
\usepackage{complexity}                
\usepackage{varioref}                  
\usepackage{hyperref}                  
\usepackage{cleveref}                  
\usepackage{tikz}                      
\usepackage[ruled,vlined]{algorithm2e} 
\usepackage{amsmath}                   
\usepackage{longtable}
\usepackage{booktabs}                  
\usepackage{pdflscape}
\usepackage{afterpage}
\usepackage{siunitx}
\usepackage{gnuplot-lua-tikz}          
\usepackage[bottom]{footmisc}

\usetikzlibrary{arrows, shadows, calc, positioning, decorations, decorations.pathmorphing,
    decorations.pathreplacing, patterns, tikzmark, fit}

\definecolor{safelightblue}{rgb}{0.65098, 0.807843, 0.890196}
\definecolor{safedarkblue}{rgb}{0.121569, 0.470588, 0.705882}
\definecolor{safelightgreen}{rgb}{0.698039, 0.87451, 0.541176}
\definecolor{safedarkgreen}{rgb}{0.2, 0.627451, 0.172549}
\definecolor{safelightorange}{rgb}{0.996078, 0.901961, 0.807843}
\definecolor{safedarkorange}{rgb}{0.54902, 0.176471, 0.015686}
\definecolor{safemediumorange}{rgb}{0.74902, 0.505882, 0.490196}
\definecolor{safeverydarkblue}{rgb}{0.007843, 0.219608, 0.345098}
\definecolor{safenearlywhite}{rgb}{0.9, 0.9, 0.9}
\definecolor{safelightgrey}{rgb}{0.7, 0.7, 0.7}

\colorlet{gp lt color 0}{safedarkblue}
\colorlet{gp lt color 1}{safelightblue}
\colorlet{gp lt color 2}{safelightorange}
\colorlet{gp lt color 3}{safedarkorange}
\colorlet{gp lt color 4}{black}
\colorlet{gp lt color 5}{safeverydarkblue}
\colorlet{gp lt color 6}{safemediumorange}

\newcommand{\Cbest}{C^\star}
\newcommand{\bounds}{\mathit{bounds}}
\newcommand{\order}{\mathit{order}}
\newcommand{\colour}{\mathit{colour}}
\newcommand{\uncoloured}{\mathit{uncoloured}}
\newcommand{\colourable}{\mathit{colourable}}
\newcommand{\expand}{\FuncSty{expand}}
\newcommand{\colourOrder}{\FuncSty{colourOrder}}
\newcommand{\vrej}{v_{\mathit{rej}}}
\newcommand{\kwunset}{\KwSty{unset}}
\newcommand{\vertexset}{V}
\newcommand{\neighbourhood}{N}

\tikzset{vertex/.style={draw, circle, inner sep=0pt, minimum size=0.5cm, font=\small\bfseries}}
\tikzset{notvertex/.style={vertex, color=white, text=black}}
\tikzset{plainvertex/.style={vertex}}
\tikzset{selectedvertex/.style={vertex, fill=safedarkblue}}
\tikzset{delvertex/.style={vertex, dotted, color=safelightgrey}}
\tikzset{vertexc1/.style={vertex, fill=safelightblue}}
\tikzset{vertexc2/.style={vertex, fill=safedarkblue, text=safenearlywhite}}
\tikzset{vertexc3/.style={vertex, fill=safelightorange}}
\tikzset{vertexc4/.style={vertex, fill=safedarkorange, text=safenearlywhite}}

\tikzset{edge/.style={color=safelightgrey}}
\tikzset{bedge/.style={ultra thick}}
\tikzset{deledge/.style={dotted, color=safelightgrey}}
\tikzset{edgel1/.style={color=safeverydarkblue}}
\tikzset{edgel2/.style={color=safemediumorange}}
\tikzset{edgel3/.style={ultra thick, color=safedarkblue}}
\tikzset{edgel4/.style={ultra thick, color=safelightorange}}

\tikzset{processarrow/.style={->, very thick, decorate, decoration={snake, post length=0.5mm}}}
\tikzset{brace/.style={decorate, decoration={brace}, very thick}}
\tikzset{label/.style={font=\small}}

\newcommand{\mclabel}[1]{\label{line:mc:#1}}
\newcommand{\mcline}[1]{line~\ref{line:mc:#1}}
\newcommand{\mclinerange}[2]{lines~\ref{line:mc:#1} to~\ref{line:mc:#2}}

\crefname{algocf}{Algorithm}{Algorithms}
\Crefname{algocf}{Algorithm}{Algorithms}
\crefname{figure}{Figure}{Figures}
\Crefname{figure}{Figure}{Figures}
\crefname{table}{Table}{Tables}
\Crefname{table}{Table}{Tables}

\makeatletter
\def\Gm@hrule{\color{black!10!white}\hrule height 0.2pt depth\z@ width\textwidth}
\def\Gm@hruled{}
\newcommand*{\gmshow@textheight}{\textheight}
\newdimen\gmshow@@textheight
\g@addto@macro\landscape{\gmshow@@textheight=\hsize\renewcommand*{\gmshow@textheight}{\gmshow@@textheight}}
\def\Gm@vrule{\color{black!10!white}\vrule width 0.2pt height\gmshow@textheight depth\z@}
\makeatother



\begin{document}

\title{Finding Maximum $k$-Cliques Faster \\ using Lazy Global Domination}

\author{
    Ciaran McCreesh%
    \thanks{This work was supported by the Engineering and Physical Sciences Research Council [grant number EP/K503058/1]}
    \\ \href{mailto:c.mccreesh.1@research.gla.ac.uk}{\nolinkurl{c.mccreesh.1@research.gla.ac.uk}}
    \and
    Patrick Prosser
    \\ \href{mailto:patrick.prosser@glasgow.ac.uk}{\nolinkurl{patrick.prosser@glasgow.ac.uk}}
}

\maketitle

\begin{abstract}
    A clique in a graph is a set of vertices, each of which is adjacent to every other vertex in
    this set. A $k$-clique relaxes this requirement, requiring vertices to be within a distance $k$
    of each other, rather than directly adjacent. In theory, a maximum clique algorithm can easily
    be adapted to solve the maximum $k$-clique problem. We use a state of the art maximum clique
    algorithm to show that this is feasible in practice, and introduce a lazy global domination rule
    which sometimes vastly reduces the search space. We include experimental results for a range of
    real-world and benchmark graphs, and a detailed look at random graphs.
\end{abstract}

\section{Introduction}

\begin{figure}[b]
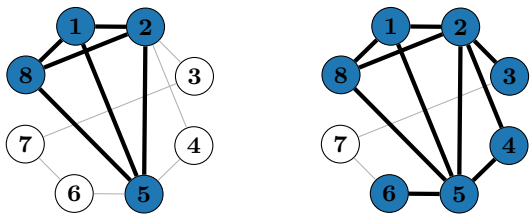
 
    \centering
    \begin{tikzpicture}
        \input{figure-k-clique-1}
    \end{tikzpicture}\hspace{4em}\begin{tikzpicture}
        \input{figure-k-clique-2}
    \end{tikzpicture}

    \caption{On the left, a graph, with its unique maximum clique $\{1, 2, 5, 8\}$ of size 4
        highlighted. On the right, the same graph, with a maximum $2$-clique $\{1, 2, 3, 4, 5, 6, 8\}$
        of size 7 highlighted. This is not a $2$-club, since the only path of length 2 between vertices 3 and 6
        goes through vertex 7. A $3$-clique covers the entire graph.}
    \label{figure:k-cliques}
\end{figure}

A clique in a graph is a set of vertices, each of which is adjacent to every other vertex in the
set. Finding a clique of maximum size in a graph is one of the basic \NP-hard
problems \cite{Garey:1990}; applications include geometry, coding theory, computer vision and
bioinformatics \cite{Bomze:1999,Butenko:2006}. However, when analysing real-world data, a clique may
be too strong a requirement. A $k$-clique (or sometimes $n$-clique or $s$-clique) is a relaxed form
of clique, where instead of requiring each pair of vertices to be directly adjacent, we only require
that they be connected by a path of length at most $k$ \cite{Luce:1950}. Thus a $1$-clique \emph{is} a
clique, a $2$-clique may be thought of as ``a group of people, all of whom either know each other or
have a mutual acquaintance'', and so on. We illustrate this in \cref{figure:k-cliques}.
Determining the size of a maximum $k$-clique is \NP-hard for any fixed $k$ \cite{Bourjolly:2002}.

A related relaxation is a $k$-club, which tightens the requirement of a $k$-clique as follows
\cite{Mokken:1979}. In a $k$-clique, each pair of vertices is connected by a path of length at most
$k$, but that path may use any vertices in the original graph. In a $k$-club, each pair of vertices
must be connected by a path of length at most $k$ using only vertices that are also in the club.
Thus the $2$-clique in \cref{figure:k-cliques} is \emph{not} a $2$-club (obviously, every $k$-club
is a $k$-clique).

A recent survey by Shahinpour and Butenko discusses algorithms and results for $k$-clique and
$k$-club problems \cite{Shahinpour:2013}.  We adopt their notation of $\tilde{\omega}_k$ for the
size of a maximum $k$-clique; the use of $\omega$ for the size of a maximum clique is standard.
They note that ``unlike the maximum clique problem, the maximum $s$-clique problem has not been the
subject of extensive research and we are not aware of any computational results for this problem to
date''. This is in contrast to the $k$-club problem, for which a wide range of computational results
are available
\cite{Bourjolly:2000,Bourjolly:2002,Mahdavi:2012,Hartung:2012,Chang:2013,Shahinpour:2013,Wotzlaw:2014}.

A maximum clique algorithm can easily be adapted to find a maximum $k$-clique in a graph $G$ by
considering the graph $G^k$, which we describe below. However, it is not obvious that this is a
viable approach: even if $G$ is sparse, $G^k$ may not be, and the maximum clique problem on dense
graphs can be very challenging computationally. Here we take a state-of-the-art maximum clique
algorithm which is suitable for use on dense graphs \cite{Prosser:2012,McCreesh:2013}, and
investigate whether this approach is feasible in practice. We modify the algorithm to include a new
lazy ``global domination'' inference step---this technique provides no benefit for typical maximum
clique problems, but for maximum $k$-clique graphs it sometimes gives improvements of several orders
of magnitude. We present computational results for the maximum $k$-clique problem on a range of
benchmark and real-world graphs. We finish with a detailed look at random graphs.

Throughout, our graphs are finite, undirected, and contain no loops. If $G$ is a graph with vertex
set $V$ and edge set $E$, we may write $\vertexset(G)$ to mean $V$. The \emph{neighbourhood} of a
vertex $v$ in a graph $G$, written $\neighbourhood_G(v)$, is the set of vertices adjacent to $v$.
The \emph{degree} of a vertex is the cardinality of its neighbourhood. The density of a graph,
denoted $D$, is the proportion of distinct pairs of vertices which have an edge between them. The
subgraph \emph{induced by} a set of vertices $W$ is the subgraph with vertex set $W$, and all edges
from the original graph that are between pairs of vertices in $W$. If $A$ and $B$ are sets of
vertices, we write $A \setminus B$ for the set of vertices which are in $A$ but not $B$, and we
write $A + v$ and $A - v$ for $A \cup \{v\}$ and $A \setminus \{v\}$ respectively.

\section{Algorithms}

Our approach for finding a maximum $k$-clique is presented as \cref{algorithm:maxKClique}. Our first
step (\mcline{power}) is to replace our input graph $G$ with a modified graph $G^k$. This graph has
the same vertex set as $G$, and edges between any two distinct vertices $v_1$ and $v_2$ iff there is
a path of length at most $k$ between $v_1$ and $v_2$ in $G$. We may construct this graph using a
bounded breadth-first search: we refer to Chang et al.\ \cite{Chang:2013} for how to implement this
quickly in practice. Now it is easy to see that maximum cliques in $G^k$ correspond with maximum
$k$-cliques in $G$ \cite{Balasundaram:2005}.

\begin{algorithm}\DontPrintSemicolon
    \begin{tikzpicture}[remember picture,overlay]
        \coordinate (cvrej1c) at ($(pic cs:cvrej1) + (0, 0.15)$);
        \coordinate (cvrej2c) at ($(pic cs:cvrej2) + (0, 0.00)$);
        \node [fill=safelightblue, rounded corners, fit=(cvrej1c) (cvrej2c)] { };

        \coordinate (cdom1c) at ($(pic cs:cdom1) + (0, 0.15)$);
        \coordinate (cdom2c) at ($(pic cs:cdom2) + (0, -0.10)$);
        \node [fill=safelightblue, rounded corners, fit=(cdom1c) (cdom2c)] { };

        \coordinate (cvinp1c) at ($(pic cs:cvinp1) + (0, 0.15)$);
        \coordinate (cvinp2c) at ($(pic cs:cvinp2) + (1, 0.00)$);
        \node [fill=safelightblue, rounded corners, fit=(cvinp1c) (cvinp2c)] { };

        \coordinate (cvrejg1c) at ($(pic cs:cvrejg1) + (0, 0.15)$);
        \coordinate (cvrejg2c) at ($(pic cs:cvrejg2) + (0, 0.00)$);
        \node [fill=safelightblue, rounded corners, fit=(cvrejg1c) (cvrejg2c)] { };

    \end{tikzpicture}

    \nl $\FuncSty{maximumKClique}$ :: (Graph $G$, Integer $k$) $\rightarrow$ Vertex Set \;
    \nl \Begin{
        \nl $G$ $\gets$ $G^k$ \mclabel{power} \;
        \nl permute $G$ so that vertices are in non-increasing degree order \mclabel{permute} \;
        \nl $\KwSty{global}$ $\Cbest$ $\gets$ $\emptyset$ \mclabel{cbest} \;
        \nl $\expand$($\emptyset$, $\vertexset(G)$) \mclabel{c} \mclabel{p} \;
        \nl $\KwSty{return}$ $\Cbest$ (unpermuted) \;
    }
    \;

    \nl $\expand$ :: (Vertex Set $C$, Vertex Set $P$) \;
    \nl \Begin{
        \nl ($\order$, $\bounds$) $\gets$ $\colourOrder$($P$) \mclabel{colouring} \;
        \nl \tikzmark{cvrej1}$\vrej$ $\gets$ $\kwunset$ \tikzmark{cvrej2} \;
        \nl \For{$i$ $\gets$ $|P|$ $\KwSty{downto}$ 1\mclabel{righttoleft}}{
            \nl \lIf{\textnormal{$|C|$ + $\bounds[i]$ $\le$ $|\Cbest|$}\mclabel{bound}}{$\KwSty{return}$}
            \nl \tikzmark{cdom1}\If{\textnormal{$\vrej$ $\ne$ $\kwunset$}}{
                \nl $P$ $\gets$ $P$ $\setminus$ $\{ w \in \vertexset(G) :
                \neighbourhood_G(w) - \vrej \subseteq \neighbourhood_G(\vrej) - w \}$
                \tikzmark{cdom2} \mclabel{dominated} \;
            }
            \nl $v$ $\gets$ $\order[i]$ \mclabel{v} \;
            \nl \tikzmark{cvinp1}\If{$v$ $\in$ $P$\tikzmark{cvinp2}\mclabel{alreadyrejected}}{
                \nl $C$ $\gets$ $C + v$ \mclabel{vinstart} \;
                \nl \lIf{\textnormal{$|C| > |\Cbest|$}}{$\Cbest$ $\gets$ $C$\mclabel{unseat}}
                \nl $P'$ $\gets$ $P$ $\cap$ $\neighbourhood_G(v)$ \mclabel{pprime} \;
                \nl \lIf{$P'$ $\ne$ $\emptyset$}{$\expand$($C$, $P'$)\mclabel{vinend}\mclabel{recurse}}
                \nl \tikzmark{creject1}$C$ $\gets$ $C - v$ \mclabel{vnotinstart} \;
                \nl $P$ $\gets$ $P - v$\tikzmark{creject2} \mclabel{vnotinend} \;
            }
            \nl \tikzmark{cvrejg1}$\vrej$ $\gets$ $v$ \tikzmark{cvrejg2} \mclabel{vrej} \;
        }
    }
    \;

    \nl $\colourOrder$ :: (Vertex Set $P$) $\rightarrow$ (Vertex Array, Int Array) \;
    \nl \Begin{
        \nl ($\order$, $\bounds$) $\gets$ ($[]$, $[]$) \;
        \nl $\uncoloured$ $\gets$ $P$ \;
        \nl $\colour$ $\gets$ $1$ \mclabel{colour1} \;
        \nl \While{$\uncoloured$ $\ne$ $\emptyset$\mclabel{uncoloured}}{
            \nl $\colourable$ $\gets$ $\uncoloured$ \mclabel{currentcolourstart} \;
            \nl \While{$\colourable$ $\ne$ $\emptyset$}{
                \nl $v$ $\gets$ the first vertex of $\colourable$ \;
                \nl append $v$ to $\order$, and append $\colour$ to $\bounds$ \;
                \nl $\uncoloured$ $\gets$ $\uncoloured - v$ \;
                \nl $\colourable$ $\gets$ $\colourable \cap \overline{\neighbourhood_G(v)}$ \mclabel{currentcolourend}\mclabel{notcolourable}\;
            }
            \nl $\colour$ $\gets$ $\colour + 1$ \mclabel{colourplus1}
        }
        \nl $\KwSty{return}$ ($\order$, $\bounds$) \;
    }

    \caption{An algorithm for the maximum $k$-clique problem.}
    \label{algorithm:maxKClique}
\end{algorithm}

\paragraph{Colouring}

The current state-of-the-art for the maximum clique problem on dense graphs, due to Tomita et al.\
\cite{Tomita:2003,Tomita:2007,Tomita:2010}, is to use branch and bound with a greedy graph
colouring.  A \emph{colouring} of a graph is an assignment of colours to vertices, such that
adjacent vertices are given different colours; if we can colour a graph using $c$ colours, then the
graph cannot contain a clique of size greater than $c$ (each vertex in a clique must be given a
different colour).

Obtaining a minimal colouring is \NP-hard, but we may create a greedy colouring in polynomial time.
This is done by the $\colourOrder$ routine: we start the first colour (\mcline{colour1}), and while
there are uncoloured vertices remaining (\mcline{uncoloured}), we try to give each vertex in turn
the current colour (\mclinerange{currentcolourstart}{currentcolourend}).  When we cannot colour any
further vertices, we start a new colour (\mcline{colourplus1}).

\begin{figure}[b]
    \centering
    \begin{tikzpicture}
        \input{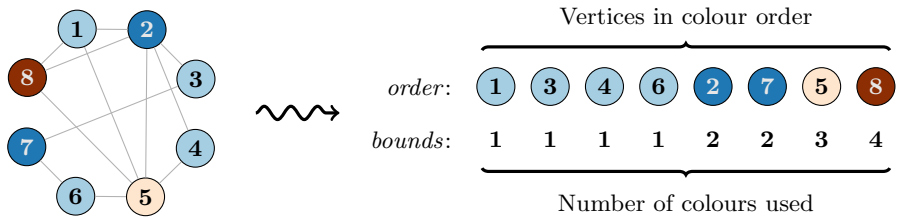}
    \end{tikzpicture}

    \caption{The graph on the left has been coloured greedily, using four colours: vertices 1, 3, 4
        then 6 were given the first colour, then vertices 2 then 7 were given the second colour,
        then vertex 5 was given the third colour, and vertex 8 the fourth colour. The $\order$ array
        contains the vertices in the order they were coloured; the $i$th entry of the $\bounds$ array
        contains the number of colours used to colour the first $i$ vertices of $\order$.}
    \label{figure:colours}
\end{figure}

The key step in Tomita's algorithms is to produce a constructive colouring, which is used in a
clever way.  The $\colourOrder$ routine does not just return the number of colours used. Instead, it
returns a pair of arrays, $\order$ and $\bounds$. The $\order$ array contains vertices, in the order
in which they were coloured. The $i$th entry of the $\bounds$ array contains the colour number used
for the $i$th vertex in $\order$. We illustrate this in \cref{figure:colours}. Crucially, $\bounds$
is non-decreasing (i.e.\ $\bounds[i+1] \ge \bounds[i]$), and we may colour the subgraph induced by
the first $i$ vertices of $\order$ using $\bounds[i]$ colours.

The order in which vertices are selected for colouring can have a large effect upon performance.
Various initial vertex orderings have been considered for the maximum clique problem---we refer to a
computational study by Prosser for details \cite{Prosser:2012}. Here we will colour vertices in a
static non-increasing degree order, which we do by permuting the graph at the top of search
(\mcline{permute}). We will \emph{not} be using Tomita et al.'s dynamic tie-breaking mechanism
\cite{Tomita:2010}: although doing so can sometimes be beneficial for small dense graphs in a
maximum clique context, for the larger graphs we will be considering here the cubic cost is
prohibitively expensive. For the same reason, we use a simple greedy colouring and do not use Tomita
et al.'s (cubic) colour repair step \cite{Tomita:2010}.

\paragraph{Branching and recursing}

We may now describe the main recursive part of the algorithm. If $v$ is a vertex, then a clique in
$G^k$ either contains only $v$ and vertices adjacent to $v$, or does not contain $v$. This allows us
to grow cliques by repeatedly picking a vertex, and branching upon whether or not to include it. Our
growing clique is stored in the variable $C$, which is initially empty (\mcline{c}). We also track
which vertices may still be added to $C$ in the variable $P$, which initially contains every vertex
(\mcline{p}). The $\expand$ procedure picks a vertex $v$ (\mcline{v}), then considers adding $v$ to
$C$ (\mclinerange{vinstart}{vinend}): we create a new $P'$ from $P$ (\mcline{pprime}) by
rejecting vertices which are not adjacent to $v$ (and thus every vertex in $P'$ is adjacent to
\emph{every} vertex in $C$). If vertices remain in $P'$, we recurse (\mcline{recurse}). We then take
the opposite branch choice, and consider rejecting from $P$ and $C$
(\mclinerange{vnotinstart}{vnotinend}). We then loop, and pick a new $v$.

\paragraph{Integrating the colour bound}

We keep track of the best solution we have found so far, which we call the \emph{incumbent}; this is
stored in $\Cbest$, which is initially empty (\mcline{cbest}). Whenever we find a new clique, we
compare its size to that of $\Cbest$, and if it is better, the incumbent is unseated
(\mcline{unseat}). Now we may make use of the colour bound. At the start of the recursive procedure
(\mcline{colouring}), we use $\colourOrder$ to produce a constructive greedy colouring of the
subgraph induced by $P$ into the array $\order$, with the colour numbers placed in $\bounds$. When
selecting $v$, we iterate over $\bounds$ from right to left (\mcline{righttoleft}). Now on
\mcline{bound} we know that the largest possible clique we could find at the current location has
size no greater than $|C| + \bounds[i]$, so if this cannot unseat the incumbent then we may abandon
search and backtrack.

\paragraph{Lazy global domination}

Aside from the $G^k$ step, what we have described so far is a standard maximum clique algorithm, and
all we have done is opted out of certain more computationally expensive inference steps (more
complicated initial vertex orderings, and cubic colourings). If we ignore the lines shown in blue,
we obtain the maximum clique algorithm variation that Prosser \cite{Prosser:2012} calls ``MCSa1''.
Now we will introduce a new lazy global domination rule which performs additional inference during
search. This rule is not specific to the maximum $k$-clique problem, and is also valid for the
maximum clique problem.

Let $v$ and $w$ be distinct vertices in a graph $G$ (they may or may not be adjacent). We say that
$v$ \emph{dominates} $w$ if the neighbourhood of $w$, excluding $v$, is a (possibly non-strict)
subset of the neighbourhood of $v$, excluding $w$. From a maximum clique perspective, this means
that $v$ is ``better than'' $w$. If $v$ and $w$ are adjacent, any clique containing $w$ may always
be extended by the inclusion of $v$; if $v$ and $w$ are non-adjacent, replacing $w$ with $v$ in any
clique containing $w$ cannot reduce the amount by which the clique may be grown.

Suppose a graph does contain one or more pairs of dominating vertices. We could make use of this
fact during search in at least two ways. Firstly, when accepting a vertex $w$, we may also
unconditionally accept any vertex $v$ which dominates $w$. Secondly, when rejecting a vertex $v$, we
may also unconditionally reject any vertex $w$ which is dominated by $v$.  We could also choose to
calculate domination globally (i.e.\ with respect to $G^k$, or even the original $G$), or locally
(i.e.\ with respect to the subgraph of $G^k$ induced by $C \cup P$).

Detecting whether one vertex dominates another may be done in linear time (we discuss this further
below), but finding all vertices dominated by a particular vertex is quadratic, and finding all
dominations is cubic. This is a heavy price to pay, if there are no dominating vertices.  This is
why such a rule has not previously been used in the maximum clique context: in the authors'
experience, most graphs typically considered for the maximum clique problem do not contain
dominating vertices, and those that do are too easy computationally for the step to be worthwhile.

However, some of the graphs we consider in the following section \emph{do} contain dominating
vertices, and although the maximum clique problem is trivial on these graphs, the maximum $k$-clique
problem is not for some values of $k$. Preliminary experiments suggested that the use of a
domination rule could be extremely beneficial in certain circumstances, but that in cases where it
had little effect, doing such a calculation introduced a substantial penalty to runtimes. Moreover,
even in graphs where dominating vertices are present, knowing this fact is sometimes not useful: it
is common for an optimal solution to be found straight away, and for the bound to be strong enough
to prove optimality immediately, so no branching occurs.

This motivates the design of a lazy global domination rule. We perform our domination checks
globally, with respect to $G^k$ (which may contain more dominating vertices than $G$), and we
remember and reuse the results of any domination checks we perform. We also only perform inference
on the ``reject'' case, to avoid introducing any cost when a solution is found and proven optimal
without branching.

The lines marked in blue in \cref{algorithm:maxKClique} show how this is done. When a vertex $\vrej$
is rejected, we remove from $P$ any vertex that is dominated (with respect to $G^k$) by $\vrej$.
This is \mcline{dominated}; the set of dominated vertices calculated here should be cached.  One
might expect that this calculation would appear after \mcline{vnotinend}. However, this introduces a
cost if the bound allows the next choice of $v$ to be eliminated. Thus we simply remember that we
have rejected $v$ by storing it in $\vrej$ (\mcline{vrej}), and lazily postpone the filtering until
after the bound has been checked.

Finally, note that we do not perform a new colouring when we reject dominated vertices---doing so
typically does not lead to a smaller bound, since most colour classes contain many vertices. Thus
when we select a $v$ from $\order$, it is now possible that $v$ has already been rejected. We check
for this on line \mcline{alreadyrejected}.

\paragraph{Bitset encoding} San Segundo et al.\ \cite{SanSegundo:2011,SanSegundo:2013} observed that
the performance of Tomita's algorithms could be enhanced substantially by using a bitset encoding to
obtain a form of SIMD-like parallelism, without altering the steps taken. We have taken such an
approach here too, although we do not describe it explicitly---when permuting $G$ on line
\mcline{permute}, the graph should be re-encoded as an array of adjacency bitsets. (It is not
helpful to do this before constructing $G^k$.) Now the intersection on \mcline{pprime} becomes a
simple bitwise ``and'' operation, and the intersection with complement on \mcline{notcolourable} is
a bitwise ``and not'' operation. This is beneficial when testing for dominance, too: each bit in the
dominated set on \mcline{dominated} may be determined by a bitwise ``and not'', unsetting a bit, and
testing whether the result is empty; the set difference is again a bitwise ``and not'' operation.

\section{Experimental Results}

Here we give experimental results on a range of standard benchmarks, and on real-world and random
graphs. Where timing results are reported, the experiments were run on a machine with Intel E5645
processors, and single-threaded runtimes are given. The time taken to read in the graph from a file
is excluded, but preprocessing time (including the construction of $G^k$ and the bitset encoding) is
included. We use the term \emph{nodes} to refer to the number of recursive calls made by the
branch-and-bound part of the algorithm.

\subsection{Real-World Graphs}

We begin with a selection of real-world and standard benchmark graphs. We look at $k$ equal to 2, 3
and 4 in every case---this is a standard practice for the $k$-club problem
\cite{Chang:2013,Wotzlaw:2014}.

\afterpage{
\setlength{\LTcapwidth}{\textwidth}
\centering
\begin{longtable}{l c r rrr rrr}
    \caption{Experimental results for a range of graphs. For each graph, we consider $k$ equal to 2,
        3 and 4. In each case we show the density of $G^k$, and then for both the unmodified
        algorithm and the algorithm with our lazy global domination step, we give the size of a
        maximum $k$-clique, the number of nodes required, and the runtime in seconds. Some results
        were aborted after one day.}
    \label{table:results}
    \\

    \toprule

    & & & \multicolumn{3}{c}{Unmodified} & \multicolumn{3}{c}{With Domination} \\
    \cmidrule(lr){4-6}
    \cmidrule(lr){7-9}
    Instance & \multicolumn{1}{c}{$k$} & \multicolumn{1}{c}{$D$} &
    \multicolumn{1}{c}{$\tilde{\omega}_k$} & \multicolumn{1}{c}{Nodes} & \multicolumn{1}{c}{Time} &
    \multicolumn{1}{c}{$\tilde{\omega}_k$} & \multicolumn{1}{c}{Nodes} & \multicolumn{1}{c}{Time} \\
    \midrule

    \endfirsthead

    \caption[]{(continued from previous page)} \\
    \toprule

    & & & \multicolumn{3}{c}{Unmodified} & \multicolumn{3}{c}{With Domination} \\
    \cmidrule(lr){4-6}
    \cmidrule(lr){7-9}
    Instance & \multicolumn{1}{c}{$k$} & \multicolumn{1}{c}{$D$} &
    \multicolumn{1}{c}{$\tilde{\omega}_k$} & \multicolumn{1}{c}{Nodes} & \multicolumn{1}{c}{Time} &
    \multicolumn{1}{c}{$\tilde{\omega}_k$} & \multicolumn{1}{c}{Nodes} & \multicolumn{1}{c}{Time} \\
    \midrule

    \endhead

    \midrule
    \endfoot

    \bottomrule
    \endlastfoot

    Erdos971
& 2
&
0.09&

42
&

42
&

0.0 &

42
&

42
&

0.0 
\\*
\hspace*{1em}\color{gray}$|V| = 472$
& 3
&
0.31&

117
&

121
&

0.0 &

117
&

119
&

0.0 
\\*
\hspace*{1em}\color{gray}$|E| = 1314$
& 4
&
0.56&

235
&

468
&

0.0 &

235
&

468
&

0.0 
\\[0.2cm]
Erdos972
& 2
&
0.01&

258
&

258
&

0.7 &

258
&

258
&

0.7 
\\*
\hspace*{1em}\color{gray}$|V| = 5488$
& 3
&
0.09&

517
&

537
&

1.0 &

517
&

521
&

1.0 
\\*
\hspace*{1em}\color{gray}$|E| = 8972$
& 4
&
0.35&

1509
&

\num{5.3e+07}&

41811.3 &

1509
&

8197
&

6.4 
\\[0.2cm]
Erdos981
& 2
&
0.09&

43
&

43
&

0.0 &

43
&

43
&

0.0 
\\*
\hspace*{1em}\color{gray}$|V| = 485$
& 3
&
0.31&

123
&

358
&

0.0 &

123
&

354
&

0.0 
\\*
\hspace*{1em}\color{gray}$|E| = 1381$
& 4
&
0.57&

245
&

246
&

0.0 &

245
&

246
&

0.0 
\\[0.2cm]
Erdos982
& 2
&
0.01&

274
&

274
&

0.8 &

274
&

274
&

0.8 
\\*
\hspace*{1em}\color{gray}$|V| = 5822$
& 3
&
0.09&

547
&

555
&

1.1 &

547
&

547
&

1.1 
\\*
\hspace*{1em}\color{gray}$|E| = 9505$
& 4
&
0.35&
\color{gray}
${\ge}1587$
&
\color{gray}
\num{1.2e+08}&
\color{gray}
1 day
&

1594
&

618826
&

424.1 
\\[0.2cm]
Erdos991
& 2
&
0.09&

43
&

44
&

0.0 &

43
&

44
&

0.0 
\\*
\hspace*{1em}\color{gray}$|V| = 492$
& 3
&
0.31&

126
&

375
&

0.0 &

126
&

374
&

0.0 
\\*
\hspace*{1em}\color{gray}$|E| = 1417$
& 4
&
0.57&

246
&

491
&

0.0 &

246
&

491
&

0.0 
\\[0.2cm]
Erdos992
& 2
&
0.01&

277
&

277
&

0.9 &

277
&

277
&

0.9 
\\*
\hspace*{1em}\color{gray}$|V| = 6100$
& 3
&
0.09&

562
&

573
&

1.2 &

562
&

562
&

1.2 
\\*
\hspace*{1em}\color{gray}$|E| = 9939$
& 4
&
0.35&
\color{gray}
${\ge}1636$
&
\color{gray}
\num{1.2e+08}&
\color{gray}
1 day
&

1643
&

202543
&

145.6 
\\[0.2cm]
Erdos02
& 2
&
0.02&

508
&

508
&

1.2 &

508
&

508
&

1.2 
\\*
\hspace*{1em}\color{gray}$|V| = 6927$
& 3
&
0.20&

1014
&

1022
&

1.9 &

1014
&

1015
&

1.9 
\\*
\hspace*{1em}\color{gray}$|E| = 11850$
& 4
&
1.00&

6927
&

6927
&

17.5 &

6927
&

6927
&

17.6 
\\

    \midrule
    c-fat200-1
& 2
&
0.13&

18
&

41
&

0.0 &

18
&

35
&

0.0 
\\*
\hspace*{1em}\color{gray}$|V| = 200$
& 3
&
0.19&

24
&

74
&

0.0 &

24
&

48
&

0.0 
\\*
\hspace*{1em}\color{gray}$|E| = 1534$
& 4
&
0.24&

30
&

134
&

0.0 &

30
&

65
&

0.0 
\\[0.2cm]
c-fat200-2
& 2
&
0.27&

35
&

35
&

0.0 &

35
&

35
&

0.0 
\\*
\hspace*{1em}\color{gray}$|V| = 200$
& 3
&
0.39&

46
&

488
&

0.0 &

46
&

102
&

0.0 
\\*
\hspace*{1em}\color{gray}$|E| = 3235$
& 4
&
0.50&

57
&

1496
&

0.0 &

57
&

128
&

0.0 
\\[0.2cm]
c-fat200-5
& 2
&
0.71&

87
&

11513
&

0.0 &

87
&

257
&

0.0 
\\*
\hspace*{1em}\color{gray}$|V| = 200$
& 3
&
1.00&

200
&

200
&

0.0 &

200
&

200
&

0.0 
\\*
\hspace*{1em}\color{gray}$|E| = 8473$
& 4
&
1.00&

200
&

200
&

0.0 &

200
&

200
&

0.0 
\\[0.2cm]
c-fat500-1
& 2
&
0.06&

21
&

52
&

0.0 &

21
&

43
&

0.0 
\\*
\hspace*{1em}\color{gray}$|V| = 500$
& 3
&
0.09&

28
&

28
&

0.0 &

28
&

28
&

0.0 
\\*
\hspace*{1em}\color{gray}$|E| = 4459$
& 4
&
0.11&

35
&

35
&

0.0 &

35
&

35
&

0.0 
\\[0.2cm]
c-fat500-2
& 2
&
0.12&

39
&

134
&

0.0 &

39
&

79
&

0.0 
\\*
\hspace*{1em}\color{gray}$|V| = 500$
& 3
&
0.17&

52
&

52
&

0.0 &

52
&

52
&

0.0 
\\*
\hspace*{1em}\color{gray}$|E| = 9139$
& 4
&
0.22&

65
&

65
&

0.0 &

65
&

65
&

0.0 
\\[0.2cm]
c-fat500-5
& 2
&
0.31&

96
&

10133
&

0.1 &

96
&

196
&

0.0 
\\*
\hspace*{1em}\color{gray}$|V| = 500$
& 3
&
0.44&

128
&

128
&

0.0 &

128
&

128
&

0.0 
\\*
\hspace*{1em}\color{gray}$|E| = 23191$
& 4
&
0.56&
\color{gray}
${\ge}155$
&
\color{gray}
\num{2.1e+10}&
\color{gray}
1 day
&

159
&

326
&

0.1 
\\[0.2cm]
c-fat500-10
& 2
&
0.62&
\color{gray}
${\ge}187$
&
\color{gray}
\num{1.6e+10}&
\color{gray}
1 day
&

189
&

560
&

0.1 
\\*
\hspace*{1em}\color{gray}$|V| = 500$
& 3
&
0.87&

252
&

252
&

0.1 &

252
&

252
&

0.1 
\\*
\hspace*{1em}\color{gray}$|E| = 46627$
& 4
&
1.00&

500
&

500
&

0.1 &

500
&

500
&

0.1 
\\[0.2cm]
p-hat300-1
& 2
&
1.00&

299
&

299
&

0.0 &

299
&

299
&

0.0 
\\*
\hspace*{1em}\color{gray}$|V| = 300$
& 3
&
1.00&

300
&

300
&

0.0 &

300
&

300
&

0.0 
\\*
\hspace*{1em}\color{gray}$|E| = 10933$
& 4
&
1.00&

300
&

300
&

0.0 &

300
&

300
&

0.0 
\\

    \midrule
    3elt
& 2
&
0.00&

10
&

340
&

0.5 &

10
&

340
&

0.7 
\\*
\hspace*{1em}\color{gray}$|V| = 4720$
& 3
&
0.01&

16
&

1582
&

0.5 &

16
&

1582
&

1.1 
\\*
\hspace*{1em}\color{gray}$|E| = 13722$
& 4
&
0.01&

27
&

911
&

0.6 &

27
&

911
&

0.8 
\\[0.2cm]
4elt
& 2
&
0.00&

11
&

486
&

5.5 &

11
&

486
&

7.5 
\\*
\hspace*{1em}\color{gray}$|V| = 15606$
& 3
&
0.00&

20
&

717
&

5.5 &

20
&

717
&

8.3 
\\*
\hspace*{1em}\color{gray}$|E| = 45878$
& 4
&
0.00&

36
&

345
&

5.6 &

36
&

345
&

6.1 
\\[0.2cm]
add20
& 2
&
0.04&

124
&

124
&

0.1 &

124
&

124
&

0.2 
\\*
\hspace*{1em}\color{gray}$|V| = 2395$
& 3
&
0.25&

671
&

671
&

0.3 &

671
&

671
&

0.3 
\\*
\hspace*{1em}\color{gray}$|E| = 7462$
& 4
&
0.67&

1454
&

1454
&

0.7 &

1454
&

1454
&

0.7 
\\[0.2cm]
add32
& 2
&
0.00&

32
&

32
&

0.6 &

32
&

32
&

0.6 
\\*
\hspace*{1em}\color{gray}$|V| = 4960$
& 3
&
0.01&

99
&

286
&

0.6 &

99
&

194
&

0.6 
\\*
\hspace*{1em}\color{gray}$|E| = 9462$
& 4
&
0.03&

268
&

268
&

0.6 &

268
&

268
&

0.6 
\\[0.2cm]
bcsstk29
& 2
&
0.01&

72
&

9752
&

5.4 &

72
&

963
&

10.5 
\\*
\hspace*{1em}\color{gray}$|V| = 13992$
& 3
&
0.02&

132
&

\num{2.7e+07}&

3141.2 &

132
&

7781
&

17.4 
\\*
\hspace*{1em}\color{gray}$|E| = 302748$
& 4
&
0.04&
\color{gray}
${\ge}204$
&
\color{gray}
\num{3.7e+08}&
\color{gray}
1 day
&

210
&

21689
&

27.8 
\\[0.2cm]
bcsstk30
& 2
&
0.01&

219
&

224
&

19.9 &

219
&

219
&

20.5 
\\*
\hspace*{1em}\color{gray}$|V| = 28924$
& 3
&
0.03&

496
&

509
&

23.3 &

496
&

496
&

24.3 
\\*
\hspace*{1em}\color{gray}$|E| = 1007284$
& 4
&
0.05&

843
&

854
&

29.2 &

843
&

845
&

30.8 
\\[0.2cm]
bcsstk31
& 2
&
0.00&

189
&

189
&

29.0 &

189
&

189
&

30.2 
\\*
\hspace*{1em}\color{gray}$|V| = 35588$
& 3
&
0.01&

278
&

605
&

30.4 &

278
&

369
&

32.0 
\\*
\hspace*{1em}\color{gray}$|E| = 572914$
& 4
&
0.02&

428
&

119640
&

202.7 &

428
&

6588
&

48.1 
\\[0.2cm]
bcsstk33
& 2
&
0.03&

141
&

141
&

2.1 &

141
&

141
&

2.1 
\\*
\hspace*{1em}\color{gray}$|V| = 8738$
& 3
&
0.08&

228
&

26033
&

10.8 &

228
&

1744
&

5.8 
\\*
\hspace*{1em}\color{gray}$|E| = 291583$
& 4
&
0.15&

435
&

\num{2.0e+06}&

825.1 &

435
&

52779
&

32.7 
\\[0.2cm]
crack
& 2
&
0.00&

10
&

2894
&

2.4 &

10
&

2894
&

10.4 
\\*
\hspace*{1em}\color{gray}$|V| = 10240$
& 3
&
0.00&

17
&

4996
&

2.5 &

17
&

4987
&

14.0 
\\*
\hspace*{1em}\color{gray}$|E| = 30380$
& 4
&
0.01&

31
&

2173
&

2.6 &

31
&

2173
&

9.3 
\\[0.2cm]
cs4
& 2
&
0.00&

6
&

5780
&

11.4 &

6
&

5780
&

97.8 
\\*
\hspace*{1em}\color{gray}$|V| = 22499$
& 3
&
0.00&

12
&

7812
&

11.8 &

12
&

7812
&

109.6 
\\*
\hspace*{1em}\color{gray}$|E| = 43858$
& 4
&
0.00&

18
&

29032
&

13.6 &

18
&

29032
&

196.3 
\\[0.2cm]
cti
& 2
&
0.00&

7
&

8918
&

6.6 &

7
&

8918
&

84.1 
\\*
\hspace*{1em}\color{gray}$|V| = 16840$
& 3
&
0.00&

15
&

6406
&

6.8 &

15
&

6406
&

60.2 
\\*
\hspace*{1em}\color{gray}$|E| = 48232$
& 4
&
0.01&

26
&

62316
&

11.7 &

26
&

62316
&

162.3 
\\[0.2cm]
data
& 2
&
0.01&

18
&

638
&

0.2 &

18
&

617
&

0.3 
\\*
\hspace*{1em}\color{gray}$|V| = 2851$
& 3
&
0.02&

32
&

4982
&

0.2 &

32
&

4913
&

0.4 
\\*
\hspace*{1em}\color{gray}$|E| = 15093$
& 4
&
0.04&

52
&

40095
&

0.8 &

52
&

36089
&

0.9 
\\[0.2cm]
fe-4elt2
& 2
&
0.00&

13
&

61
&

2.8 &

13
&

61
&

3.0 
\\*
\hspace*{1em}\color{gray}$|V| = 11143$
& 3
&
0.00&

20
&

389
&

2.8 &

20
&

389
&

4.7 
\\*
\hspace*{1em}\color{gray}$|E| = 32818$
& 4
&
0.01&

32
&

448
&

2.9 &

32
&

446
&

4.6 
\\[0.2cm]
fe-pwt
& 2
&
0.00&

16
&

95
&

29.6 &

16
&

95
&

34.4 
\\*
\hspace*{1em}\color{gray}$|V| = 36519$
& 3
&
0.00&

29
&

167
&

30.6 &

29
&

167
&

33.7 
\\*
\hspace*{1em}\color{gray}$|E| = 144794$
& 4
&
0.00&

52
&

224
&

29.7 &

52
&

224
&

32.9 
\\[0.2cm]
fe-sphere
& 2
&
0.00&

7
&

14173
&

6.4 &

7
&

14173
&

106.2 
\\*
\hspace*{1em}\color{gray}$|V| = 16386$
& 3
&
0.00&

12
&

34328
&

7.3 &

12
&

34328
&

205.8 
\\*
\hspace*{1em}\color{gray}$|E| = 49152$
& 4
&
0.00&

19
&

73632
&

10.1 &

19
&

73632
&

283.5 
\\[0.2cm]
memplus
& 2
&
0.02&

574
&

574
&

7.9 &

574
&

574
&

7.8 
\\*
\hspace*{1em}\color{gray}$|V| = 17758$
& 3
&
0.26&

8057
&

8061
&

74.7 &

8057
&

8058
&

80.1 
\\*
\hspace*{1em}\color{gray}$|E| = 54196$
& 4
&
0.74&

8963
&

8963
&

106.0 &

8963
&

8963
&

112.9 
\\[0.2cm]
uk
& 2
&
0.00&

5
&

433
&

0.5 &

5
&

433
&

0.9 
\\*
\hspace*{1em}\color{gray}$|V| = 4824$
& 3
&
0.00&

8
&

1891
&

0.5 &

8
&

1891
&

1.4 
\\*
\hspace*{1em}\color{gray}$|E| = 6837$
& 4
&
0.01&

14
&

2168
&

0.5 &

14
&

2168
&

1.1 
\\[0.2cm]
vibrobox
& 2
&
0.02&

121
&

302
&

4.0 &

121
&

302
&

5.1 
\\*
\hspace*{1em}\color{gray}$|V| = 12328$
& 3
&
0.08&

408
&

1984
&

7.5 &

408
&

1984
&

12.0 
\\*
\hspace*{1em}\color{gray}$|E| = 165250$
& 4
&
0.26&
\color{gray}
${\ge}1094$
&
\color{gray}
\num{8.8e+07}&
\color{gray}
1 day
&
\color{gray}
${\ge}1094$
&
\color{gray}
\num{8.7e+07}&
\color{gray}
1 day

\\[0.2cm]
whitaker3
& 2
&
0.00&

9
&

1222
&

2.3 &

9
&

1222
&

7.3 
\\*
\hspace*{1em}\color{gray}$|V| = 9800$
& 3
&
0.00&

15
&

3724
&

2.3 &

15
&

3724
&

12.2 
\\*
\hspace*{1em}\color{gray}$|E| = 28989$
& 4
&
0.01&

23
&

6530
&

2.4 &

23
&

6530
&

14.8 
\\[0.2cm]
wing-nodal
& 2
&
0.01&

29
&

648
&

3.1 &

29
&

648
&

5.3 
\\*
\hspace*{1em}\color{gray}$|V| = 10937$
& 3
&
0.02&

54
&

13091
&

5.1 &

54
&

13039
&

23.6 
\\*
\hspace*{1em}\color{gray}$|E| = 75488$
& 4
&
0.04&

114
&

\num{6.0e+07}&

4639.0 &

114
&

\num{6.0e+07}&

4676.0 
\\

    \midrule
    adjnoun
& 2
&
0.50&

50
&

50
&

0.0 &

50
&

50
&

0.0 
\\*
\hspace*{1em}\color{gray}$|V| = 112$
& 3
&
0.91&

83
&

164
&

0.0 &

83
&

164
&

0.0 
\\*
\hspace*{1em}\color{gray}$|E| = 425$
& 4
&
0.99&

107
&

107
&

0.0 &

107
&

107
&

0.0 
\\[0.2cm]
as-22july06
& 2
&
0.04&

2391
&

2391
&

18.6 &

2391
&

2391
&

19.1 
\\*
\hspace*{1em}\color{gray}$|V| = 22963$
& 3
&
0.36&

8455
&

673880
&

13356.2 &

8455
&

94497
&

1834.0 
\\*
\hspace*{1em}\color{gray}$|E| = 48436$
& 4
&
0.79&

14911
&

14911
&

286.3 &

14911
&

14911
&

288.8 
\\[0.2cm]
astro-ph
& 2
&
0.01&

361
&

365
&

7.1 &

361
&

362
&

7.1 
\\*
\hspace*{1em}\color{gray}$|V| = 16706$
& 3
&
0.10&

1553
&

1567
&

16.8 &

1553
&

1560
&

20.1 
\\*
\hspace*{1em}\color{gray}$|E| = 121251$
& 4
&
0.35&
\color{gray}
${\ge}4040$
&
\color{gray}
\num{1.2e+07}&
\color{gray}
1 day
&
\color{gray}
${\ge}4040$
&
\color{gray}
\num{1.1e+07}&
\color{gray}
1 day

\\[0.2cm]
celegans-meta
& 2
&
0.44&

238
&

238
&

0.0 &

238
&

238
&

0.0 
\\*
\hspace*{1em}\color{gray}$|V| = 453$
& 3
&
0.89&

371
&

371
&

0.0 &

371
&

371
&

0.0 
\\*
\hspace*{1em}\color{gray}$|E| = 2025$
& 4
&
0.98&

432
&

432
&

0.0 &

432
&

432
&

0.0 
\\[0.2cm]
celegansneural
& 2
&
0.55&

135
&

135
&

0.0 &

135
&

135
&

0.0 
\\*
\hspace*{1em}\color{gray}$|V| = 297$
& 3
&
0.95&

245
&

245
&

0.0 &

245
&

245
&

0.0 
\\*
\hspace*{1em}\color{gray}$|E| = 2148$
& 4
&
1.00&

295
&

295
&

0.0 &

295
&

295
&

0.0 
\\[0.2cm]
cond-mat
& 2
&
0.00&

108
&

108
&

6.5 &

108
&

108
&

6.8 
\\*
\hspace*{1em}\color{gray}$|V| = 16726$
& 3
&
0.01&

250
&

1403
&

7.8 &

250
&

844
&

7.7 
\\*
\hspace*{1em}\color{gray}$|E| = 47594$
& 4
&
0.05&
\color{gray}
${\ge}649$
&
\color{gray}
\num{7.3e+07}&
\color{gray}
1 day
&

720
&

674453
&

823.3 
\\[0.2cm]
cond-mat-2003
& 2
&
0.00&

203
&

204
&

22.7 &

203
&

204
&

22.6 
\\*
\hspace*{1em}\color{gray}$|V| = 31163$
& 3
&
0.02&
\color{gray}
${\ge}629$
&
\color{gray}
\num{6.5e+07}&
\color{gray}
1 day
&
\color{gray}
${\ge}629$
&
\color{gray}
\num{6.2e+07}&
\color{gray}
1 day

\\*
\hspace*{1em}\color{gray}$|E| = 120029$
& 4
&
0.12&
\color{gray}
${\ge}2605$
&
\color{gray}
\num{1.8e+07}&
\color{gray}
1 day
&
\color{gray}
${\ge}2606$
&
\color{gray}
\num{1.8e+07}&
\color{gray}
1 day

\\[0.2cm]
cond-mat-2005
& 2
&
0.00&

279
&

279
&

37.8 &

279
&

279
&

39.0 
\\*
\hspace*{1em}\color{gray}$|V| = 40421$
& 3
&
0.03&
\color{gray}
${\ge}1060$
&
\color{gray}
\num{2.0e+07}&
\color{gray}
1 day
&
\color{gray}
${\ge}1060$
&
\color{gray}
\num{1.9e+07}&
\color{gray}
1 day

\\*
\hspace*{1em}\color{gray}$|E| = 175691$
& 4
&
0.16&
\color{gray}
${\ge}4185$
&
\color{gray}
\num{6.2e+06}&
\color{gray}
1 day
&
\color{gray}
${\ge}4185$
&
\color{gray}
\num{6.1e+06}&
\color{gray}
1 day

\\[0.2cm]
dolphins
& 2
&
0.32&

14
&

14
&

0.0 &

14
&

14
&

0.0 
\\*
\hspace*{1em}\color{gray}$|V| = 62$
& 3
&
0.59&

30
&

30
&

0.0 &

30
&

30
&

0.0 
\\*
\hspace*{1em}\color{gray}$|E| = 159$
& 4
&
0.77&

40
&

40
&

0.0 &

40
&

40
&

0.0 
\\[0.2cm]
email
& 2
&
0.09&

72
&

72
&

0.0 &

72
&

72
&

0.0 
\\*
\hspace*{1em}\color{gray}$|V| = 1133$
& 3
&
0.45&

233
&

19031
&

0.7 &

233
&

19031
&

0.7 
\\*
\hspace*{1em}\color{gray}$|E| = 5451$
& 4
&
0.86&

654
&

6854
&

0.5 &

654
&

6676
&

0.5 
\\[0.2cm]
football
& 2
&
0.45&

17
&

147
&

0.0 &

17
&

145
&

0.0 
\\*
\hspace*{1em}\color{gray}$|V| = 115$
& 3
&
0.95&

69
&

70
&

0.0 &

69
&

70
&

0.0 
\\*
\hspace*{1em}\color{gray}$|E| = 613$
& 4
&
1.00&

115
&

115
&

0.0 &

115
&

115
&

0.0 
\\[0.2cm]
hep-th
& 2
&
0.00&

51
&

51
&

1.6 &

51
&

51
&

1.6 
\\*
\hspace*{1em}\color{gray}$|V| = 8361$
& 3
&
0.01&

125
&

239
&

1.7 &

125
&

176
&

1.8 
\\*
\hspace*{1em}\color{gray}$|E| = 15751$
& 4
&
0.04&

347
&

158164
&

49.6 &

347
&

23714
&

8.6 
\\[0.2cm]
jazz
& 2
&
0.69&

103
&

107
&

0.0 &

103
&

107
&

0.0 
\\*
\hspace*{1em}\color{gray}$|V| = 198$
& 3
&
0.95&

174
&

174
&

0.0 &

174
&

174
&

0.0 
\\*
\hspace*{1em}\color{gray}$|E| = 2742$
& 4
&
0.99&

192
&

192
&

0.0 &

192
&

192
&

0.0 
\\[0.2cm]
karate
& 2
&
0.61&

18
&

18
&

0.0 &

18
&

18
&

0.0 
\\*
\hspace*{1em}\color{gray}$|V| = 34$
& 3
&
0.86&

25
&

25
&

0.0 &

25
&

25
&

0.0 
\\*
\hspace*{1em}\color{gray}$|E| = 78$
& 4
&
0.99&

33
&

33
&

0.0 &

33
&

33
&

0.0 
\\[0.2cm]
lesmis
& 2
&
0.43&

37
&

37
&

0.0 &

37
&

37
&

0.0 
\\*
\hspace*{1em}\color{gray}$|V| = 77$
& 3
&
0.85&

58
&

58
&

0.0 &

58
&

58
&

0.0 
\\*
\hspace*{1em}\color{gray}$|E| = 254$
& 4
&
0.99&

75
&

75
&

0.0 &

75
&

75
&

0.0 
\\[0.2cm]
netscience
& 2
&
0.01&

35
&

35
&

0.1 &

35
&

35
&

0.1 
\\*
\hspace*{1em}\color{gray}$|V| = 1589$
& 3
&
0.01&

54
&

54
&

0.1 &

54
&

54
&

0.1 
\\*
\hspace*{1em}\color{gray}$|E| = 2742$
& 4
&
0.02&

85
&

85
&

0.1 &

85
&

85
&

0.1 
\\[0.2cm]
PGPgiantcompo
& 2
&
0.00&

206
&

206
&

2.6 &

206
&

206
&

2.7 
\\*
\hspace*{1em}\color{gray}$|V| = 10680$
& 3
&
0.02&

423
&

843
&

3.1 &

423
&

841
&

3.1 
\\*
\hspace*{1em}\color{gray}$|E| = 24316$
& 4
&
0.07&

1161
&

1161
&

4.7 &

1161
&

1161
&

4.8 
\\[0.2cm]
polblogs
& 2
&
0.27&

352
&

352
&

0.1 &

352
&

352
&

0.1 
\\*
\hspace*{1em}\color{gray}$|V| = 1490$
& 3
&
0.58&

776
&

2210
&

0.6 &

776
&

2177
&

0.6 
\\*
\hspace*{1em}\color{gray}$|E| = 16715$
& 4
&
0.66&

1127
&

1537
&

0.8 &

1127
&

1166
&

0.7 
\\[0.2cm]
polbooks
& 2
&
0.37&

28
&

28
&

0.0 &

28
&

28
&

0.0 
\\*
\hspace*{1em}\color{gray}$|V| = 105$
& 3
&
0.64&

54
&

54
&

0.0 &

54
&

54
&

0.0 
\\*
\hspace*{1em}\color{gray}$|E| = 441$
& 4
&
0.86&

68
&

68
&

0.0 &

68
&

68
&

0.0 
\\[0.2cm]
power
& 2
&
0.00&

20
&

20
&

0.6 &

20
&

20
&

0.6 
\\*
\hspace*{1em}\color{gray}$|V| = 4941$
& 3
&
0.00&

30
&

30
&

0.6 &

30
&

30
&

0.6 
\\*
\hspace*{1em}\color{gray}$|E| = 6594$
& 4
&
0.01&

61
&

61
&

0.6 &

61
&

61
&

0.6 
\\

\end{longtable}}

\paragraph{Erd\H{o}s collaboration graphs}

In the first part of \cref{table:results} we present experimental results from Erd\H{o}s
collaboration graphs from the Pajek dataset by Vladimir Batagelj and Andrej
Mrvar\footnote{\url{http://vlado.fmf.uni-lj.si/pub/networks/data/}}.  We were able to solve all of
these problems in under eight minutes (and all but three in under two seconds) when using the
domination rule. However, using the unmodified maximum clique algorithm, two of these results did
not finish running within one day. Note that for $k = 4$, a $k$-clique covers all of ``Erdos02''.

In several cases, the algorithm found and proved an optimal solution immediately ($\tilde{\omega}_k$
is equal to the number of search nodes). This illustrates the necessity of laziness: if we simply
computed dominating pairs upfront, we would be paying a cubic preprocessing cost for an algorithm
which is effectively quadratic in practice.

By comparing these results with the $k$-club results of Chang et al.\ \cite{Chang:2013}, we see that
in all but four cases the $k$-clique and $k$-club numbers are equal; all of these differences occur
when $k = 4$. (Chang et al.\ did not investigate the ``Erdos02'' graph, but Wotzlaw
\cite{Wotzlaw:2014} confirmed privately that the $k$-clique and $k$-club numbers are the same here
too.) On the other hand, the $k$-clique numbers are sometimes much easier to find, both
algorithmically and computationally.

\paragraph{Clique graphs}

In the second part of \cref{table:results} we present results from the ``clique'' graphs from the
Second DIMACS implementation challenge\footnote{\url{http://dimacs.rutgers.edu/Challenges/}}. These
graphs were designed to test maximum clique implementations. Nearly all of these graphs have
diameter 2, so a 2-clique covers the entire graph---we have ignored these. The only exceptions are
the ``c-fat'' family (all of which are trivial for a maximum clique solver), and one of the
``p\_hat'' graphs.

With the domination rule, we solve all of these problems within a tenth of a second. Without, two of
the results take over a day, and the rest remain trivial. Note that in several cases, for some
values of $k$ a $k$-clique covers the entire graph.  Again using Chang et al.'s results
\cite{Chang:2013}, we see that for the first six graphs in this table the $k$-clique and $k$-club
numbers are the same for each value of $k$ (Chang et al.\ did not investigate ``c-fat500-10'' or
``p-hat300-1'').

\paragraph{Partitioning graphs}

The third part of \cref{table:results} presents results from the smallest 20 partitioning graphs
from the 10th DIMACS Implementation
Challenge\footnote{\url{http://staffweb.cms.gre.ac.uk/~wc06/partition/}}.  Many of these graphs are
considerably larger than those typically considered for the maximum clique problem, and we might
expect our $O(|V|^2)$ memory requirements to cause problems. Nonetheless, with the domination rule
there is only one instance which we were unable to solve within a day (and without the domination
rule, there are two).

On the other hand, we sometimes see a significant cost where the domination rule does not help, and
where the proof of optimality is not immediate: in ``3elt'' and ``4elt'', our runtimes can nearly
double, and for ``cs4'' and ``cti'' the slowdown is sometimes over a factor of ten. In other words,
laziness does not help when the rule turns out to be used, but useless.

Five of these graphs were considered for the $k$-club problem by Wotzlaw \cite{Wotzlaw:2014}. In all
five cases, the $k$-clique and $k$-club numbers are the same for $k$ equal to 2, 3 and 4. However,
the $k$-clique number was again consistently much easier to find.

\paragraph{Clustering graphs}

The final part of \cref{table:results} presents results from the smallest 20 partitioning graphs
from the 10th DIMACS Implementation
Challenge\footnote{\url{http://www.cc.gatech.edu/dimacs10/archive/clustering.shtml}}. Again, from a
maximum clique perspective these would be considered unusually large graphs. However, only five were
unsolvable within a day (plus a sixth when the domination rule was not used), and half of the
problems took under two seconds.

Seven of these graphs were considered for the $k$-club problem by Wotzlaw \cite{Wotzlaw:2014}. In
these cases, the $2$-clique and $2$-club numbers are the same, except for ``football'' where the
$2$-club number is 16 but the $2$-clique number is 17; for $k = 3$ and $k = 4$ there are some
differences. The difference in computational difficulty between the $k$-clique and $k$-club problems
really stands out here: for ``polblogs'' with $k = 3$ and $k = 4$, Wotzlaw was unable to prove
optimality within an hour, but we required less than a second to do so. In both of these cases the
$k$-clique and $k$-club numbers are the same, which suggests a potential improvement to $k$-club
algorithms: first solve the maximum $k$-clique problem instead, and test whether the result found is
a $k$-club, before embarking upon a more complicated search. (Note that a negative result does not
imply that the $k$-clique and $k$-club numbers necessarily differ, since solutions are not unique.)

\subsection{Random Graphs}

An Erd\H{o}s-R\'{e}nyi random graph $G(n, p)$ has $n$ vertices, and an edge between each distinct
pair of vertices with probability $p$, chosen independently. Here we investigate the size of a
maximum $k$-clique in such graphs, and the complexity of finding it. In each case, we use an average
over 100 samples for every point.  We do not use the domination rule for these experiments: the
probability of random graphs having dominating vertices is very low.

In \cref{figure:graph-omega} we illustrate the average value of $\tilde{\omega}_k$ in $G(200,
p)$ for different values of $k$, and a range of values of $p$ for the $x$-axis. We see that even for
very low edge probabilities, a maximum $k$-clique quickly covers the entire graph.  (This is in contrast to
the maximum clique problem, where a maximum clique does not even cover a quarter of the graph for
edge probabilities below 0.75.) In \cref{figure:graph-nodes} we show the average size of the
search space (number of nodes, or recursive calls made) for the same problem. We see that there is a
complexity peak for each $k$, although the peak is much smaller for $k = 4$ than it is for $k = 3$,
which is in turn much smaller than it is for $k = 2$. The peak also occurs for lower edge
probabilities as $k$ increases. For contrast, for the maximum clique problem, the peak occurs at
around edge probability 0.9, and is two orders of magnitude larger.

\begin{figure}[p] 
    \centering
    \input{gen-graph-omega}
    \caption{Values of $\tilde{\omega}_k$ for random graphs $G(200, p)$, with varying edge probabilities.
        We see that even for very low edge probabilities, a
    maximum $k$-clique quickly covers the entire graph. This is in contrast to maximum cliques, which remain
    small even at much higher edge probabilities.}
    \label{figure:graph-omega}

    \vspace{1.5em}

    \input{gen-graph-nodes}
    \caption{Search space size for random graphs $G(200, p)^k$, with varying edge probabilities. We see that
        $4$-clique is easier than $3$-clique in practice, which in turn is easier than $2$-clique. (The
        complexity peak for maximum clique occurs at around edge probability 0.9, and requires
        approximately 15 million search nodes.)}
    \label{figure:graph-nodes}
\end{figure}

In \cref{figure:graph-omega-2,figure:graph-nodes-2} we show the effect of
changing $n$ and fixing $k = 2$. As $n$ increases from 50 to 200, the complexity peak becomes much
more pronounced, and shifts slightly towards the left (lower edge probabilities).

\begin{figure}[p] 
    \centering
    \input{gen-graph-omega-2}
    \caption{The size of a maximum $2$-clique in random graphs $G(n, p)$ with varying edge
        probabilities, and different values of $n$. For $G(50, p)$, a $2$-clique has size average 50
    from $p = 0.42$ onwards.}
    \label{figure:graph-omega-2}

    \vspace{1.5em}

    \input{gen-graph-nodes-2}
    \caption{The search space size for the maximum $2$-clique problem in random graphs $G(n, p)$
        with varying edge probabilities, and for different values of $n$. As $n$ increases, the
    complexity peak grows and moves slowly to the left.}
    \label{figure:graph-nodes-2}
\end{figure}

\section{Conclusion}

We have shown that using a maximum clique algorithm to solve the maximum $k$-clique algorithm for a
graph $G$ by considering $G^k$ in place of $G$ is feasible in practice. This is despite $G^k$
potentially being dense even if $G$ is sparse.

We introduced a new lazy global domination rule. This was sometimes extremely beneficial---without
this rule, we would have been unable to solve six of the problem instances we considered, and many
others would have taken much longer. However, even with laziness there is still sometimes a cost to
pay when this rule does nothing. This rule is thus harmful for the graphs typically considered for
the maximum clique problem, and we see the benefit of tailoring algorithms to the problem being
solved. We suggest that a similar rule may also be useful for the maximum $k$-club problem.

Quite often, we saw $k$-clique numbers and $k$-club numbers being the same. However, solving the
maximum $k$-clique problem is much easier, both in terms of the algorithm and computationally. Thus
it is worth checking whether the simpler model would be sufficient for practical applications before
trying to solve the $k$-club problem.

In random graphs, we saw that $G(n, p)^k$ is easier than $G(n, p')$ with some higher probability
$p'$. We also saw that as $k$ increases, the problem gets easier---this was not typically the case
for some of the real world graphs.

Our results suggest that $k$ is a very coarse grained parameter. We saw that often a $2$-clique or
$3$-clique would cover the entire graph. In these circumstances the increased restrictions for
$k$-club are of no benefit. It is not obvious if somehow allowing a ``fractional'' value of $k$
could give more fine-grained control. Thus it may be worth considering other clique relaxations not
based upon distance (although other models also have problems: a density-based relaxation known as
quasi-clique, for example, can allow vertices with only a single edge to be added to a ``clique''
\cite{Abello:2002}).

\bibliographystyle{amsalpha}
\bibliography{k-clique}

\end{document}